\documentclass[review]{elsarticle}

\usepackage{lineno}
\usepackage{geometry}
\usepackage{array}
\usepackage{amsmath}
\usepackage[linesnumbered]{algorithm2e}
\usepackage{lscape}
\SetKwRepeat{Do}{do}{while}

\journal{Journal of \LaTeX\ Templates}

\begin{document}
\begin{frontmatter}
\title{Variable neighborhood search for partitioning sparse biological networks into the maximum edge-weighted $k$-plexes}
\author{Milana Grbi\'c}
\ead{milana.grbic@pmf.unibl.org}

\author{Aleksandar Kartelj}
 \ead{aleksandar.kartelj@gmail.com}
  \author{Savka Jankovi\'c}
 \ead{savka.jankovic@pmf.unibl.org}
\author{Dragan Mati\'c}
 \ead{dragan.matic@pmf.unibl.org}
  \author{Vladimir Filipovi\'c}
 \ead{vladofilipovic@hotmail.com}

\begin{abstract}
In a network, a $k$-plex represents a subset of $n$ vertices where the degree of each vertex in the subnetwork induced by this subset is at least $n-k$. The maximum edge-weight $k$-plex partitioning problem (Max-EkPP) is to find the $k$-plex partitioning in edge-weighted network, such that the sum of edge weights is maximal. The Max-EkPP has an important role in discovering new information in large sparse biological networks. We propose a variable neighborhood search (VNS) algorithm for solving Max-EkPP. The VNS implements a local search based on the 1-swap first improvement strategy and the objective function that takes into account the degree of every vertex in each partition. The objective function favors feasible solutions, also enabling a gradual increase in terms of objective function value when moving from slightly infeasible to barely feasible solutions. A comprehensive experimental computation is performed on real metabolic networks and other benchmark instances from literature. Comparing to the integer linear programming method from  literature, our approach succeeds to find all known optimal solutions. For all other instances, the VNS either reaches previous best known solution or improves it. The proposed VNS is also tested on a large-scaled dataset which was not previously considered in literature.
\end{abstract}


\end{frontmatter}

\section{Introduction}

In recent years there is an increasing effort to provide algorithms for better understanding of biological structures and processes. Among many other approaches, partitioning large biological networks into smaller clusters or functional modules is a commonly used technique for discovering new properties and functionalities of a particular structure.
In this work, we deal with the partitioning of the edge-weighted networks into k-plex components, where a subset of some $n$ vertices in a network is a $k$-plex if the degree of each vertex in the subnetwork induced by this subset is at least $n-k$. The aim of the maximum edge-weight $k$-plex partitioning problem (Max-EkPP) is to find the $k$-plex partitioning with the maximal total weight of edges.

Partitioning networks into high density subnetworks, especially cliques, has already been proven as a useful technique for obtaining new information in understanding complicated relations between biological elements. For example, partitioning in protein threading analysis - constrained threading problem can be reduced on maximum  edge weight clique  problem \cite{akutsu2006dynamic}. The protein side chain packing problem  is transformed into a problem of finding a maximum weight clique. The edge weighting function is defined in a way that reflects the frequency of contact pairs in a database of proteins \cite{jb2006multiple}. Finding  cliques is also one of the methods for identification of the clusters that are later divided into protein complexes and dynamic functional modules. By analyzing the multibody structure of the network of protein--protein interactions (PPI), molecular modules that are densely connected within themselves, but sparsely connected with the rest of the network, are discovered \cite{spirin2003protein}. Cliques have a similar use in  modular decomposition of PPI networks. This decomposition allows to combine proteins into the actual functional complexes by identifying groups of proteins acting as a single unit \cite{gagneur2004modular}.

  On the other hand, a number of  biological networks classes contain only sparse networks. Dealing with such networks, partitioning into cliques can be too restrictive method, so many potentially useful information about the interference of biological objects can be neglected. Therefore, clique relaxation approaches could be even more useful.
In the approach presented in this work, partitioning is followed by the principle that the objects in each partition are still highly connected in a particular way, but not so restrictively to form a clique. By relaxing  cliques to sparse graphs, biological objects become connected in semantically or functionally logical groups which we call $k$-plexes,  having in mind that the total sum of weights in all partitions should be as large as possible.

\subsection{Problem definition}
Let a network be denoted as $G = (V,E)$, where $V=\{1,2,...,n\}$ is the set of nodes and $E\subset V\times V$ is the set of edges. With $uv$ we simply denote the edge $\{u,v\}\in E$. With real numbers $w_{uv}>0$ we denote the weight of the edge connecting nodes $u$ and $v$. We call $u$ and $v$ the end-vertices of the edge $uv$.

 Let $k\geq 1$ be an integer. A set of nodes $S$ is called $k$-plex  if the degree of each node in the sub-network induced by $S$ is at least $n-k$. The weight of a $k$-plex is the sum of all its edge weights. The weight of the whole partition is the sum of the weights of all its $k$-plex components. The maximum k-plex partitioning (Max-EkPP) problem is then defined as finding such a partition of $G$ which is of the maximum total weight and each component is a $k$-plex.
If $k = 1$, the $k$-plex is a clique and the Max-EkPP is brought down to the maximum edge-weight clique partitioning problem (Max-ECP).
\subsection{Our contributions}
The contributions of this paper can be summarized as follows:

 We constructed the first heuristic method  for partitioning graph into densely connected components - $k$-plexes. The proposed method is based on the variable neighborhood search (VNS) metaheuristic.

 The proposed VNS method implements a newly created objective function which takes into consideration the degree of every vertex in each $k$-plex. Objective function favors feasible solutions over infeasible ones as expected, but also enables a gradual increase in terms of objective function value when moving from slightly infeasible to barely feasible solutions.

 We successfully applied the proposed VNS on some biological instances, as well as on some graph instances used in similar NP-hard problems.

 The quality of the proposed method is proven by the fact that it achieves all previously known optimal solutions. It also proposes new lower bounds for the instances for which the optimal solutions are unknown.

 Based on the obtained computational results and graphical interpretation of the solutions, we established a biological interpretation. This indicates the potential of using the proposed method in discovering new biological information of a particular structure.

The remainder of the paper is organized as follows. In the next section, we review some related researches. In Section  \ref{sec:vns} we describe the proposed VNS method.  We present the computational results, comparison with an existing method and biological evaluation of the obtained data in Section \ref{results}. The last section concludes the paper and suggests the future work.

\section{Literature review}

A $k$-plex structure was introduced in an early work \cite{seidman1978graph}  in late 1970s, as a clique-like structure of variable strength.The $k$-plex structure is defined as a graph with $n$ vertices, where each vertex is connected by at least $n- k$ other vertices. The optimization problem which has arisen involves the identification of the $k$-plex of the maximum cardinality in an unweighted sparse graph. The problem is called the maximum $k$-plex problem (Max-kP problem).

 Although one could expect that this formulation was immediately tackled by researchers, the problem has not been thoroughly analyzed  for more than 30 years.
Meanwhile, a progress in the Internet and other computer-based technologies, including bioinformatics, triggered the generation of tremendous amounts of various interaction data. Balasundaram et al. \cite{balasundaram2011clique} brought the mentioned problem again into the attention of the scientific community, by recognizing its close connection with behavior of some real-world networks, particularly with social networks. In the mentioned paper, the problem of identification of a maximum cardinality k-plex in an unweighted sparse graph (Max-kP) has been proven to be NP hard and an integer programming formulation (ILP) has been presented. ILP formulations developed for Maximum clique problem in \cite{martins2010extended} can be adopted for solving other related problems, including maximum size $k$-plex problem. Beside these exact methods based on the integer linear programming approach, there are some heuristic methods for solving Max-kP in the unweighted sparse graphs. For instance, McClosky and Hicks \cite{mcclosky2012combinatorial} adapted  combinatorial clique algorithms to find maximum k-plexes and proposed a new upper bound on the cardinality of k-plexes. Moser et al. \cite{moser2009algorithms} proposed some practical algorithms for finding maximum k-plexes which outperforms other approaches. $k$-plex clustering is also a way of non-hierarchical decomposition of the graph into clusters, which enables an application of parallelization algorithms.

 Several other variants of clique relaxations, as well as adequate mathematical programming formulations have been studied in \cite{pattillo2012clique}.
In large biological networks, for example PPI network, proteins with similar GO annotation can be clustered together  by partitioning biological networks into highly connected components \cite{huffner2014partitioning}.  Clustering large data sets plays an important role in gene expression analysis. In \cite{hartuv2000algorithm}, cluster analysis of cDNA fingerprints is used to identify clones corresponding to the same gene. In \cite{navlakha2010exploring}, many near-optimal clusterings are used to explore the dynamics of network clusterings. This is later applied on several biological and other networks. In order to show the types of insights that can be extracted from large collections of near-optimal solutions, the authors analyzed the ERK1/ERK2 mitogen-activated protein kinase (MAPK18) signal-transduction pathway and a network of cortical-cortical connections in the human brain.

Identifying cohesive subgroups (not necessarily cliques and $k$-plexes)  has also been performed in a number of non-biological networks: in studying terrorist and other criminal networks \cite{chen2004crime}, web graphs \cite{terveen1999constructing}, wireless networks \cite{krishna1997cluster}, in finding structural patterns embedded within social network data \cite{mukherjee2004graph}, text mining \cite{balasundaram2008cohesive},  stock markets \cite{boginski2014network}, etc.

A complementary problem to identifying k-plexes in the graph $G$ is a problem of identifying co-k-plexes in $\overline{G}$. A subset $S$ of a graph $G$ is a co-k-plex if the subgraph induced by $S$ has a maximum degree of $k-1$ or less. From this definition, one can conclude that $S$ is a co-k-plex in $G$ if and only if $S$
is a k-plex in the complement graph $\overline{G}$. For $k = 1$, we get that the co-k-plex is exactly the independent
set. Such a relaxation of the maximum clique problem is also in a close connection with defective coloring problem \cite{cowen1997defective,trukhanov2013algorithms}, which is a relaxation of the well known vertex coloring problem in a graph. A $(\kappa, d)$--coloring of a graph is a coloring of the vertices with $\kappa$ colors such that  no vertex is adjacent to more than $d$ vertices of its same color. For $d=0$, $(\kappa, d)$--coloring is the proper vertex coloring problem. For a given number $d$, identifying appropriate co-$d-1$-plexes corresponds to the $(\kappa, d)$- coloring, where vertices in a co-$d-1$-plex are allowed to be colored with the same color.

In the context of partitioning weighted graphs into distinct components,  in the literature one can find strong results addressing the maximum edge-weight cliques partitioning (Max-ECP) problem. In Max-ECP, the objective is to cluster all the vertices into disjoint cliques, such that the total sum of the edge weights of all partitions is as large as possible. Although Max-ECP is a special case of Max-EkPP for $k=1$, it has been considered  on complete graphs, both in several earlier works \cite{grotschel1989cutting,dorndorf1994fast,oosten2001clique,wang2006solving}, as well as in recent state-of-the-art proposed heuristic methods \cite{zhou2016three,brimberg2017solving}.
From the other hand, partitioning of sparse graph into cliques could be too restrictive, since many useful information regarding relations between the elements can be lost.  Following that consideration, Martins \cite{martins2016modeling} proposed a polynomial size integer linear programming formulation for  Max-EkPP problem, also considering the inclusion of additional topological constraints in the model. The performance of the proposed ILP model has been tested on biological and artificial networks, which we also used in our paper.

\section{Variable neighborhood search for Max-EkPP}\label{sec:vns}

Variable Neighborhood Search (VNS) algorithm is a robust metaheuristic introduced by Mladenovi\'c and Hansen \cite{mla97}. The main searching principle of a VNS is based on the empirical evidences: (a) multiple local optima are correlated in some sense (usually close to each other) and (b) a local optimum found in one neighborhood structure is not necessarily a local optimum for some other neighborhood structure.

\begin{figure}[h!]
\centering
\includegraphics[scale=0.7]{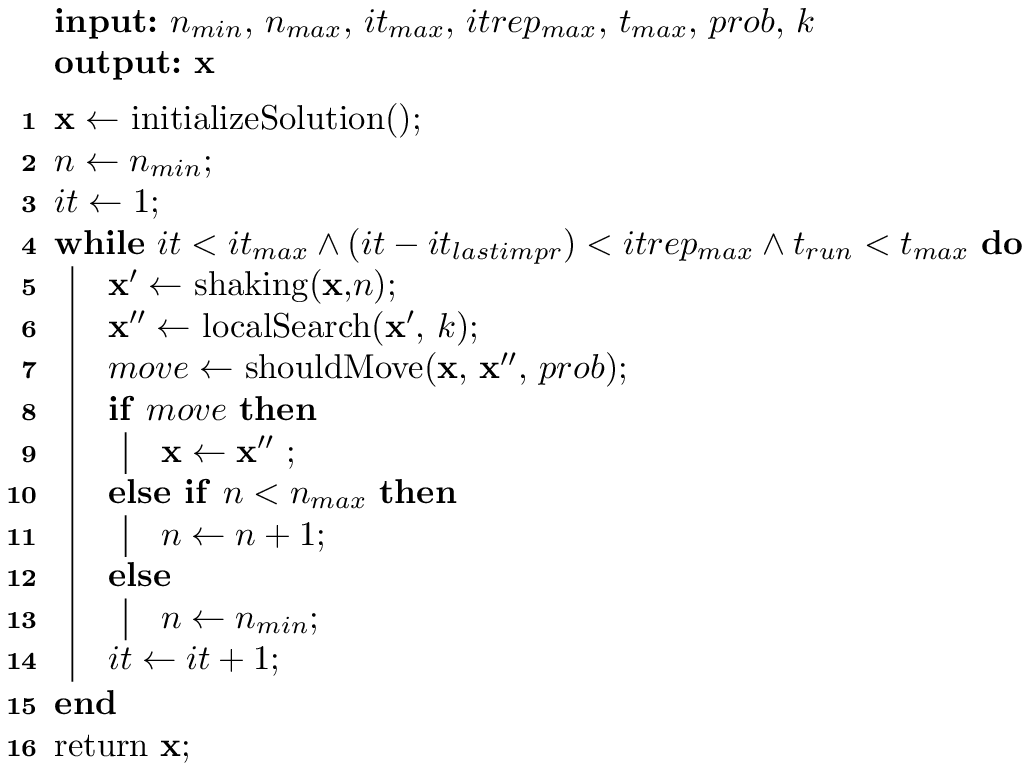}
 \caption{The overall structure of the VNS}
\label{vns}
\end{figure}

The overall structure of the VNS algorithm for Max-kP is shown on the Figure \ref{vns}.

 The input of the VNS algorithm, beside the given input graph $G$, consists of:
\begin{itemize}
\item $n_{min}$ and $n_{max}$ - minimal and maximal VNS neighborhood structure size;
\item $it_{max}$, $itrep_{max}$, $t_{max}$ - maximal number of total iterations , maximal number of iterations without improvement, and maximal execution time in seconds, respectively;
\item $prob$ - probability to move to the other solution of the same quality;
\item $k$ - integer value that corresponds to the value of k inside term Max-kP.
\end{itemize}

VNS algorithm usually imposes two main procedures: shaking and local search (LS). Shaking procedure manages the system of the neighborhoods and in each iteration it randomly selects a new point from
the current neighborhood. The main purpose of the shaking procedure is to resolve situations when local search procedure is stuck into a local suboptimal solution. More details about shaking are given in Section \ref{shaking}.

Inside the LS procedure, the algorithm is trying to improve the solution selected by the shaking. LS systematically checks other solutions in its nearest neighborhood. Details about LS are given in Section \ref{localsearch}.

In the main loop of the algorithm, shaking procedure is iteratively called, until no further improvements of the
best solution can be made inside the current neighborhood.
When that situation appears,
the algorithm steps into the next neighborhood. When the last neighborhood $n_{max}$ is explored, the search restarts at the first neighborhood $n_{min}$ .

The execution of the VNS is stopped when either of the following conditions becomes satisfied: a maximum number of iterations is reached, a maximum number of iterations without any improvement of the current best solution is reached or a maximum allowed execution time is reached.

\subsection{Solution representation and the objective function}

Let $G=(V,E)$ be an edge-weighted network and let  $k$ be the given integer number. Recall that the aim of the Max-EkPP is to find the $k$-plex partition with the maximum total weight.

  The solution of the proposed VNS algorithm is represented by an integer array $\mathbf{x}$ of the length $|V|$. Each element of the array corresponds to one vertex of the graph, denoting to which component (partition) the corresponding vertex is assigned. More precisely, the vertex  $i$ is assigned to the component $V_j$ if $x_i=j$.
The initial solution is created by assigning each element of the array $\mathbf{x}$ random integer number from interval $[1,2,...,\sqrt{|V|}]$. The upper bound for the initial number of partitions $\sqrt{|V|}$ was empirically determined.

Unfeasible solutions are not implicitly disallowed by the representation. This is fortunate, since an objective function can direct the search to the more promising and feasible areas in a subtle way, without posing heavy penalty on a slightly infeasible solution. In other words, we constructed  the objective function with two aims: to subtly avoid infeasible solutions and to maximize the total weight of the partition.

Let $(V_1,V_2,...,V_l)$ be a (not necessarily feasible) solution of the Max-EkPP. Let $w_{total}$ be the total sum of the weights of all edges in the network $G$, i.e. $w_{total} =\sum_{uv\in E}w_{uv}$.

 We introduce the term ``correct vertex'' in a solution. A vertex $v \in V_j$, $j\in \{1,2,...,l\}$ is ``correct'' if the degree of $v$ in the network induced by $V_j$ is at least $|V_j|-k$. This means if each vertex in a partition is ``correct'', then that partition is a $k$-plex. Let $correct_{total}$ be the total number of correct vertices in the solution.

In the partition $(V_1,V_2,...,V_l)$, with $w_{sol}$ we denote the total sum of the weights of all edges with correct end vertices.

Now we can define the objective function of the solution.

\begin{equation}\label{eqn:objective}
obj(V_1,V_2,...,V_l) = correct_{total} +\frac{w_{sol}}{w_{total}}
\end{equation}
Since the value $w_{sol}/w_{total}$ is always less or equal to 1, the objective function mainly depends on the first term, i.e. the total number of correct vertices in the solution. Consequently, the objective function of any feasible solution will be greater than the objective function of any infeasible solution. If two solutions have the same number of correct vertices, then the solution with greater  total sum of the weights has also greater objective value.  As a consequence, the maximization process  discards  solutions with many incorrect vertices and directs the search into the feasible regions.


At the same time, the proposed objective function properly orders infeasible solutions. This gives better infeasible solutions higher chance to appear, which can consequently, after local search, transform them into higher quality feasible solutions.

\subsection{Shaking}\label{shaking}
The main purpose of the shaking procedure is to extend the search space of the current solution in order to reduce the possibility that the algorithm falls into suboptimal solutions.

Inside the shaking procedure, the algorithm creates a system of neighborhoods used for deriving new solutions based on the current best solution $\mathbf{x}$.

For defining the $\kappa$-th neighborhood we use the following procedure:
some $\kappa$ vertices  from $V$ are  randomly chosen.
For each chosen vertex $v$, the
algorithm changes its component as follows.

If $l$ is the total number of partitions, then an integer $q$ is randomly chosen from the set $\{1,2,...,l+1\}$. If $q<l+1$, then the vertex $v$ is moved to the existing partition $V_q$. If $q=l+1$, then a new singleton partition is established ($V_{l+1}=\{v\}$) and the total number of partitions is increased by one. If the old partition, from which the vertex $v$ was chosen, becomes empty, then the total number of partitions is decreased by one.
 This strategy allows that the total number of partitions can be changed during the searching process.

Therefore, the aim of the proposed shaking procedure is twofold: it perturb vertices from one component to another and possibly increases or decreases the total number of partitions. The solution $\mathbf{x'}$, obtained by the shaking is the subject of the further improvements in the local search.

\subsection{Local search}\label{localsearch}

\begin{figure}[h]
\centering
\includegraphics[scale=0.7]{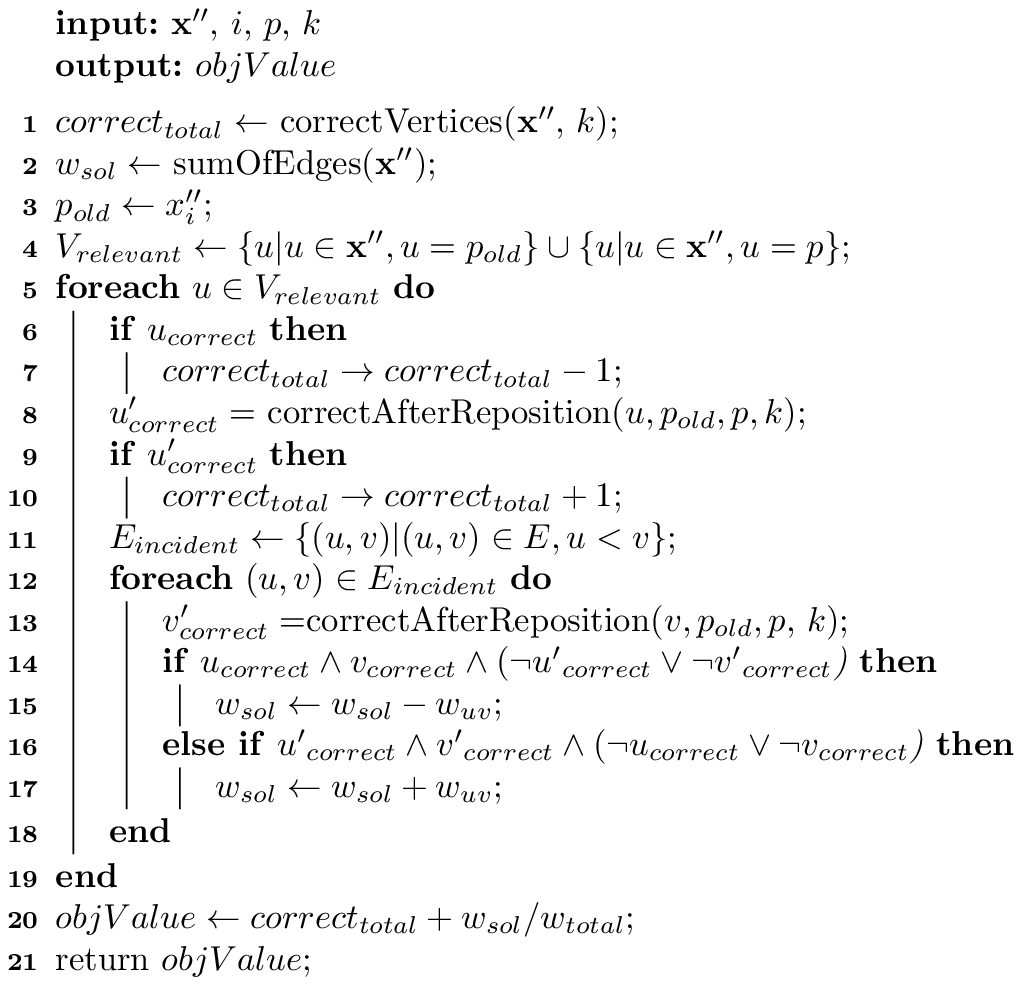}
 \caption{Partial calculation of the objective function}
 \label{figObjectiveValue}
\end{figure}

The purpose of the local search procedure is to explore  the neighborhood of a new solution space obtained through shaking in order to achieve a locally optimal solution.
\begin{figure}[h!]
\centering
\includegraphics[scale=0.7]{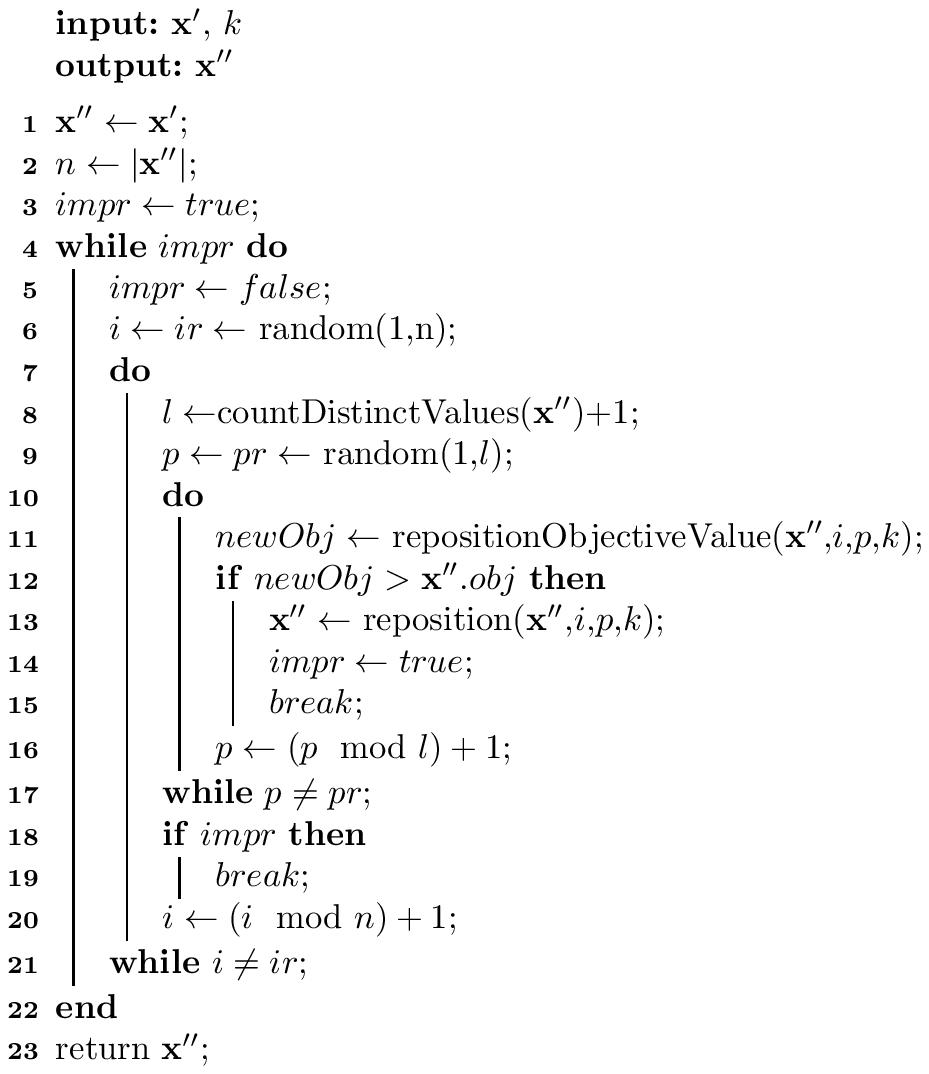}
 \caption{Local search}
 \label{figLocalSearch}
\end{figure}
The proposed LS is based on the so called ``1-swap first improvement'' strategy (see Figure \ref{figLocalSearch}).
Let $\mathbf{x'}$ be a solution suggested by the shaking procedure and let $\mathbf{x''}$ be a solution which LS is afterwards applied on.
The LS iteratively examines new solutions formed by moving a single vertex from its belonging component to some other component, in the following way. Let $v$ be a vertex which is the subject of movement. If $l$ is the total number of components, then a random integer $p$ is chosen from $\{1,2,...,l+1\}$.  If $p<l+1$, the vertex $v$ is moved to the existing partition $V_p$, else a new partition $V_{p}=\{v\}$ is established.
 By this movement step, $\mathbf{x''}$ is adjusted in accordance to the partial calculation of the objective function (Figure~\ref{figObjectiveValue}).
When the first improvement of the solution is found, the change is immediately applied and $\mathbf{x''}$ is updated. After the improvement, LS restarts the outer loop, i.e., it tries to find new improvement by resetting the initial values of the selected vertex and candidate partition.
If an improvement is still not found, LS continues with the next vertex, repeating the procedure. When the set of all vertices is exhausted without improvement, LS stops.

Since LS is usually the most time-consuming part in the entire VNS, it is of crucial importance to construct the LS procedure to be as much efficient as possible, considering also the quality of the obtained local optimum.
Local search systematically checks neighborhood of the given solution by moving one vertex from its starting partition to some other partition. This infers only a slight difference in the underlying structure of the solution so it might be useful to partially calculate objective function of the newly formed solution. Figure~\ref{figObjectiveValue} shows the pseudocode of this partial calculation.
 Current value of the objective function is composed of two terms: $correct_{total}$ represents the number of vertices that are correct, while $w_{sol}$ is the sum of edges connecting correct vertices. Partial objective function calculates the new objective function value after moving the $i^{th}$ vertex from its current partition to $p^{th}$ partition. This movement induces the change in local solution structure since only vertices from the initial and target partition, and their related edges should be considered in the partial calculation. The set of these relevant vertices is denoted by $V_{relevant}$. For each vertex $u$ in the set $V_{relevant}$, value of $correct_{total}$ is adjusted. For example, if $u$ was previously correct and now it is not correct, $correct_{total}$ is decreased by one.
Similarly, all edges incident with relevant vertices are checked, and consequently, value of $w_{sol}$ is adjusted. The adjustment takes place in two situations: firstly, if edge became correct after movement, while it was not correct before movement, and vice versa, if it is not correct after movement while it was correct previously.
The total objective function is afterward assembled from these two partially calculated terms.
If the solution cannot be further improved inside the local search, the (hopefully improved) solution is returned back to the main VNS algorithm where it is named $\mathbf{x''}$.

The next step of the algorithm is to compare the quality of the current best solution $\mathbf{x}$ to the solution $\mathbf{x''}$, obtained after shaking and LS is finished. If the value of the objective of the solution $\mathbf{x''}$ is greater than of the solution $\mathbf{x}$, then $\mathbf{x''}$ becomes the new current best solution ($\mathbf{x}=\mathbf{x''}$). If the objective function value of the solution $\mathbf{x''}$ is less than the objective function value of the solution $\mathbf{x}$, then the solution $\mathbf{x}$ remains the current best one. If the values of the objective function of both solutions are the same, then $\mathbf{x}$ is set to $\mathbf{x''}$ with a probability of $prob$.



\section{Experimental results}\label{results}
In this section we evaluate the performance of the proposed VNS method.
All experiments are performed on the Intel Xeon E5410 CPU @2.33 GHz with 16 GB RAM and Windows Server 2012 2R 64Bit operating system. For each execution only one thread/processor is used. The VNS is implemented in C programming language and
compiled with Visual Studio 2015 compiler.

In order to make the comparison to the other method from the literature as fair as possible, we used the same benchmark data sets as in \cite{martins2016modeling} and tested them for three values of $k$, namely $k \in \{1, 2, 3\}$. The first two sets contain biological instances created on metabolic reactions from \cite{forster2003genome}, as it is described in \cite{martins2016modeling}. In Table \ref{systematized} we show these biological data in a systematized view, with shorten instance names.  The third set of instances was taken from the well known DIMACS database, available at http://www.dcs.gla.ac.uk/$\sim$pat/maxClique. Testing the VNS on DIMACS instances consists of two phases. In the first phase, we followed the approach from \cite{martins2016modeling}, and took  DIMACS instances with less than 100 vertices and larger sparse instances with less than 200 vertices and density at most 0.25. In the second phase, we tested our VNS on the rest of 73 DIMACS instances.

 Since the original DIMACS instances are not weighted, like in other papers \cite{martins2016modeling,gouveia2015solving}, we also followed the weighting strategy proposed in \cite{pullan2008approximating}, setting $w_{i,j} = ((i + j) \mod 200) + 1$.
For each instance, we performed 10 independent executions of the VNS algorithm. Termination criterion
is based on the combination of three criteria: the maximum total number of iterations reached, where $it_{max} = 20000$,  the maximum number of iterations without improvement, which is set to 10000, or maximum total execution time reached, which is set to 1 hour. Other control parameters are set as follows:  $n_{min}=1$ and $n_{max} = 80$ and $prob = 0.1$.

\subsection{Experimental results obtained on SC-NIP-m-tr instances}
This section reports the experimental results for the set of  SC-NIP-m-tr, $r=1,...,5$ instances. As it is already mentioned, we considered three values of $k$, i.e. $k \in \{1,2,3\}$. Table \ref{results1} columns are organized as follows. In the first two columns, values of $k$ and a shorten name of the instance are shown. The next two columns contain the optimal value and the best known value, if the optimal value is not known; the following five columns contain data related to the VNS: the best and average VNS results obtained in 10 runs, average gap, total execution time (in seconds). If the best achieved VNS result is equal to the known optimal value, then the column $VNS_{best}$ contains the mark $opt$. If the optimal value is not known, but VNS succeeds to find the best known value, then that column contains the mark $best$. The mark $new$ is used if there are no previous results for the considered instance. The last two columns are related to the exact ILP method proposed in \cite{martins2016modeling}: column ILP contains the result obtained by the mentioned ILP, with the marks $opt$ or $best$ if the result obtained by the ILP is optimal or the best known. In cases when ILP could not find any solution due to the memory limits, the mark ``--'' is used.
\begin{table}[h!]
\begin{center}

\caption[skip=0]{A view of considered biological metabolite networks}
\begin{tabular}{@{\hspace{0.5em}}l@{\hspace{0.5em}}l@{\hspace{0.5em}}ll|@{\hspace{0.5em}}l@{\hspace{0.5em}}l@{\hspace{0.5em}}ll}
inst.    & $|V|$  & $|E|$   & density   &inst.   & $|V|$  & $|E|$   & density     \\\hline
m-t1 & 991  & 4161  & 0.0085 &r-t1 & 1393 & 56276 & 0.0580\\
m-t2 & 602  & 1520  & 0.0084 &r-t2 & 1183 & 17776 & 0.0254\\
m-t3 & 177  & 269   & 0.0173 &r-t3 & 663  & 1782  & 0.0081\\
m-t4 & 129  & 166   & 0.0201 &r-t4 & 377  & 321   & 0.0045\\
m-t5 & 75   & 84    & 0.0303  &r-t5 & 45   & 27    & 0.0272\\\hline

\end{tabular}\label{systematized}
\end{center}
\end{table}
Table \ref{results1} shows that the proposed VNS succeeds to find all 9 known optimal solutions. In addition, for each of these instances, the VNS reaches the optimal value in each of 10 runs. For the rest of 6 instances, where ILP model from \cite{martins2016modeling} could not find any solution, the VNS succeeds to find solution in a reasonable time up to 1 hour. The average gap is rather small and it is less than 1 for all instances. Thus, one can conclude that the VNS is rather stable while solving this class of instances.

\begin{table}[ht!]\caption{Experimental results obtained on SC-NIP-m-tr instances}
\scriptsize
\begin{center}
\begin{tabular}{@{\hspace{0.5em}}l@{\hspace{0.5em}}l@{\hspace{0.5em}}r@{\hspace{0.5em}}r@{\hspace{0.5em}}r@{\hspace{0.5em}}r@{\hspace{0.5em}}r@{\hspace{0.5em}}r@{\hspace{0.5em}}r@{\hspace{0.5em}}r@{\hspace{0.5em}}}
\hline
$k$&$inst.$&$opt$&$best$&$V_{best}$&$V_{avg}$&$V_{gap}$&$V_t^{tot}$&$ILP$&$ILP_t$\\
\hline
1&m-t1&1866&1866&$opt$&1864&0.11&3600.22&$opt$&2296.94\\
1&m-t2&1538&1538&$opt$&1538&0&1072.51&$opt$&1.25\\
1&m-t3&910&910&$opt$&910&0&92.96&$opt$&0.02\\
1&m-t4&831&831&$opt$&831&0&45.5&$opt$&0\\
1&m-t5&723&723&$opt$&723&0&15.73&$opt$&0\\
2&m-t1&-&2151&$new$&2147.3&0.17&3600.14&-&-\\
2&m-t2&-&1773&$new$&1771.8&0.07&1495.49&-&-\\
2&m-t3&1021&1021&$opt$&1021&0&100.74&$opt$&50.43\\
2&m-t4&907&907&$opt$&907&0&54.75&$opt$&3.03\\
2&m-t5&801&801&$opt$&801&0&16.42&$opt$&0.2\\
3&m-t1&-&2353&$new$&2337.1&0.68&3600.18&-&-\\
3&m-t2&-&1943&$new$&1939.4&0.19&1988.38&-&-\\
3&m-t3&-&1141&$new$&1141&0&121.08&-&-\\
3&m-t4&-&1022&$new$&1022&0&69.79&-&-\\
3&m-t5&887&887&$opt$&887&0&17.62&$opt$&34.2\\
\hline
\end{tabular}\label{results1}
\end{center}
\end{table}
The ILP method from \cite{martins2016modeling} was more successful for $k=1$ comparing to the two other values of $k$. It succeeded to find all optimal solutions for $k=1$, three optima for $k=2$ and one optimum for $k=1$. The proposed VNS achieves all these optimal solutions, but also succeeds to find solutions regardless of the value $k$ in a very reasonable time, up to 3020 seconds.
\subsection{Experimental results obtained on SC-NIP-r-tr instances}

This section provides experimental results obtained on the second class of biological data, i.e. SC-NIP-r-tr instances. The results are shown in Table \ref{results2}, which is organized in similar way as Table \ref{results1}. As it can be seen from Table \ref{results2}, VNS achieves all 7 known optimal solutions. For the rest of 8 instances VNS achieves the new best results. With respect to the solution quality, the picture is similar as in the case of SC-NIP-m-tr instances. The ILP method from \cite{martins2016modeling} succeeds to find 7 optimal solutions: four optima for $k=1$, two optima for $k=2$ and one optimum for $k=3$. The proposed VNS succeeds to find all these optima, also providing high quality solutions for other cases.

From computational point of view, the SC-NIP-r-tr instances are more challenging than SC-NIP-m-tr because of their dimensions (see Table \ref{systematized}). Therefore, necessary runtime is proportionally greater comparing to the execution times for SC-NIP-m-tr instances. For five SC-NIP-r-tr instances, the algorithm stopped after the time limit is reached (1 hour), while for other instances, the termination happened after maximum number of iteration was reached. The average gap for this class is again rather small (less than 1) for all instances.

\begin{table}[t!]\caption{Experimental results obtained on SC-NIP-r-tr instances}
\scriptsize
\begin{center}
\begin{tabular}{@{\hspace{0.5em}}l@{\hspace{0.5em}}l@{\hspace{0.5em}}r@{\hspace{0.5em}}r@{\hspace{0.5em}}r@{\hspace{0.5em}}r@{\hspace{0.5em}}r@{\hspace{0.5em}}r@{\hspace{0.5em}}r@{\hspace{0.5em}}r@{\hspace{0.5em}}}
\hline
$k$&$inst.$&$opt$&$best$&$V_{best}$&$V_{avg}$&$V_{gap}$&$V_t^{tot}$&$ILP$&$ILP_t$\\
\hline
1&r-t1&-&57681&$new$&57544.6&0.24&3607.77&-&\\
1&r-t2&34576&34576&$opt$&34561.6&0.04&3601.2&$opt$&4.26\\
1&r-t3&5411&5411&$opt$&5411&0&1550.95&$opt$&0.08\\
1&r-t4&1232&1232&$opt$&1232&0&327.82&$opt$&0\\
1&r-t5&140&140&$opt$&140&0&3.71&$opt$&0.02\\
2&r-t1&-&57729&$new$&57496&0.4&3602.58&-&\\
2&r-t2&-&34592&$new$&34563.6&0.08&3601.65&-&\\
2&r-t3&-&5423&$new$&5423&0&1569.11&3183&$>$10800\\
2&r-t4&1245&1245&$opt$&1245&0&331.75&$opt$&6.4\\
2&r-t5&140&140&$opt$&140&0&3.82&$opt$&0.01\\
3&r-t1&-&57775&$new$&57587.4&0.33&3602.19&-&\\
3&r-t2&-&34641&$new$&34572.5&0.2&3601.26&-&\\
3&r-t3&-&5465&$new$&5465&0&1496.84&-&\\
3&r-t4&-&1245&$new$&1245&0&327.45&-&\\
3&r-t5&140&140&$opt$&140&0&3.84&$opt$&0.14\\
\hline
\end{tabular}\label{results2}
\end{center}
\end{table}

Tables \ref{results1} and \ref{results2} show that in both classes of biological instances, execution time depends on the graph density, i.e. smaller density induces smaller execution time. A natural explanation is that smaller number of edges causes the lowering of the total number of executions of the local search procedure, which further leads to the shorter overall execution time. By comparing the values in columns $opt$ and $best$ for the same instance and different values of $k$, we conclude that the value of objective functions increases with increasing the value of $k$. This is because the total number of edges included in clusters increases with the relaxation of the adjacency conditions in each cluster.

\section{Visualization and biological explanation of the obtained results}

In the following consideration, we will show how the relaxation of the clustering requirements can lead to more useful information from biological point of view. Among many metabolic processes that appeared in various $k$-plexes obtained by the proposed VNS algorithm, we chose to discuss following processes: amino acid degradation process, fatty acids synthesis, vitamin B6 synthesis, oxidation of the succinate to the fumarate and formaldehyde oxidation.

In order to confirm the reliability of the obtained results, particular information of the biochemical pathways of considered organism Saccharomyces cerevisiae are checked and confirmed with the data presented in Yeast Pathways Database \cite{yeastDB}.

\subsection{Amino acid degradation process}\label{sec:aminoacid}
\begin{figure}[t]\begin{center}
\includegraphics[scale=0.55]{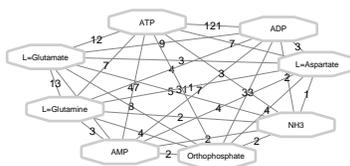}
\caption{The largest cluster obtained for $k=1$}\label{k1g1}
\end{center}
\end{figure}

In this subsection, we consider the amino acid degradation process, which is one of the most important processes in metabolism.
In Figure \ref{k1g1} we show the largest cluster obtained for $k=1$, which contains the rough representation of the amino acid degradation.

Ammonia presented in the organism is used as a resource  of nitrogen for amino-acid synthesis and if it released in larger quantity, it must be removed because of its toxicity.

 In the considered organism Saccharomyces cerevisiae, ammonia can be incorporated into the amino group of glutamate, by two pathways: the reductive amination of 2-ketoglutarate, catalyzed by glutamate dehydrogenase where NADPH serves as the source of electrons, or by the ATP-dependent synthesis of glutamine from glutamate and ammonia catalyzed by glutamine synthetase \cite{magasanik2003ammonia}.

\begin{figure}[ht!]\begin{center}
\includegraphics[scale=0.6]{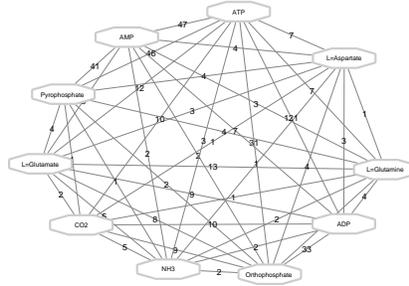}
\caption{The largest cluster obtained for $k=2$}\label{k2g1}
\end{center}
\end{figure}

This cluster is shown in Figure \ref{k1g1} is a clique with 8 vertices and contains the main intermediates which figure in  ammonia synthesis from glutamic and aspartic acids. Glutamate binds the orthophosphoric group from ATP, resulting in glutamine (ADP is formed, and orthophosphate is released).
 Further, in Figure \ref{k2g1} we show the largest cluster obtained for $k=2$. A wider set of intermediates is now shown, also including additional reactions. We again see that glutamate appears in the reaction of glutamine and L-Aspartate by ATP consumption. With this reaction asparagine is formed, which  is converted to aspartate by deamination, while the ammonia is released. Also, glutamate and ammonia is released by deamination of glutamine. If the system is extended by two additional intermediates, we see one more way of glutamate synthesis, that is reaction of $\textnormal{CO}_2$ and glutamine by ATP consumption.

\begin{figure}[ht!]\begin{center}
\includegraphics[scale=0.6]{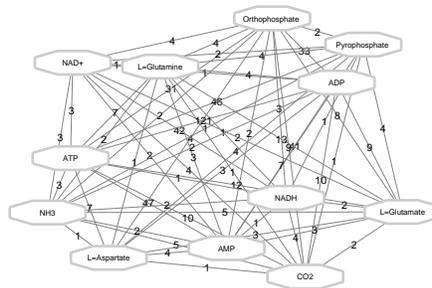}
\caption{The largest cluster obtained for $k=3$}\label{k3g1}
\end{center}
\end{figure}

More detailed graphical interpretation is shown in Figure \ref{k3g1}, obtained by the proposed algorithm for $k=3$. Since the condition for forming clusters is now more relaxed, more intermediates figures in the cluster. In addition to the previous ones, in the cluster shown in Figure \ref{k3g1} we see the process of the oxidative deamination. The oxidative deamination which occurs in cells involves the amino acid glutamate. Glutamate is oxidatively deaminated by the enzyme glutamate dehydrogenase, using NAD or NADP as a coenzyme. By this process, two toxic products are synthesized: hydrogen peroxide and ammonia. In Figure \ref{k3g1} we see that the algorithm grouped all these intermediates in one cluster, which was not the case with the more strict conditions (cases $k=1$ and $k=2$).

\subsection{Fatty acids synthesis}
\begin{figure}[ht!]
\begin{center}
\includegraphics[scale=0.45]{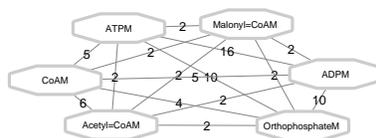}
\caption{Fatty acids synthesis cluster for $k=1$}\label{k1g2}
\end{center}
\end{figure}

\begin{figure}[t]\begin{center}
\includegraphics[scale=0.5]{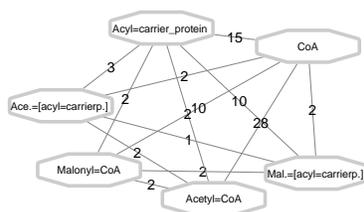}
\caption{Extended display of fatty acids synthesis for $k=2$}\label{k2g4}
\end{center}
\end{figure}

\begin{figure}[t]\begin{center}
\includegraphics[scale=0.5]{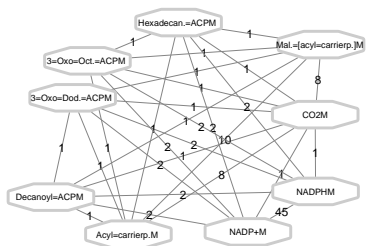}
\caption{More detailed display of fatty acids synthesis, obtained for $k=3$}\label{k3g2}
\end{center}
\end{figure}

In Figure \ref{k1g2} we show the second largest cluster obtained for $k=1$. We can see that the algorithm grouped intermediates which figure in fatty acids synthesis. Fatty acids are long molecules and the process of their synthesis can be divided in three phases. The first phase includes synthesis of the coenzyme malonil-CoA  from acetil-CoA, since the malonil-CoA is more reactive molecule and more suitable for the extension of the fatty acid chain. Acetil-CoA is synthesized from CoA by consumption of the ATP which is converted to ADP with the release of orthophosphate. As it is shown in Figure \ref{k1g2}, the edge connecting acetil-CoA and malonil-CoA is of the weight 2 since their reaction is reversible, so their link is counted twice. The second phase consists of five successive cycle reactions, starting with the binding of the acetil-CoA and malonil-CoA directly to the carrier protein, after malonil-acyl carrier protein (ACP) is formed and CoA is released. In Figure \ref{k2g4} are shown these reactions of binding amino acid with acyl carrier protein (ACP). From this figure one can see that the largest number of reactions are related to synthesis of acetil-CoA from CoA and  to binding of the malonil-CoA to carrier protein, which proves that the algorithm recognized the overall system of the biosynthesis. The reaction is continued by extending the chain by 2C atoms (new malonil-CoA is binded), until a long
 fatty acid is synthesized.
 Each time when malonile-CoA is binded with ACP, CoA is released. During the condensation with ACP, $\textnormal{CO}_2$ is removed and oxo compounds appear, which are reduced by the presence of NADPH (which is transformed to NADP+) and are hydrolyzed to enoyl compounds, which again are reduced with NADPH.
In the third phase, saturated extended fatty acid accepts new malonil-CoA and further extends the chain by the same scheme (Figure \ref{k3g2}). To conclude, the algorithm recognized the complex process by identifying the highly weighted clusters related to the fatty acids synthesis. Like in the case of amino acid degradation process explained in Section \ref{sec:aminoacid}, the results obtained by the relaxation of the condition, are used to obtain useful information of the metabolic reactions.

\begin{figure}[h]\begin{center}
\includegraphics[scale=0.45]{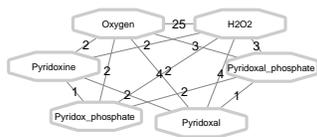}
\caption{The main intermediates for vitamin B6 synthesis, the cluster obtained for $k=2$}\label{bvitam}
\end{center}
\end{figure}

\subsection{Other useful findings}

Among other clusters obtained by the proposed algorithm on the considered network, we notice two interesting clusters obtained for $k=2$: vitamin B6 synthesis and glutathione dependent formaldehyde oxidation.
In Figure \ref{bvitam}, obtained for $k=2$ the main intermediates for vitamin B6 synthesis are shown. Pyridoxal phosphate (PLP) is the active form of vitamin B6 and is a cofactor in many reactions of amino acid metabolism \cite{yeastDB}. From the figure, we see that the algorithm grouped different forms of vitamin B6: Pyridoxine,   pyridoxal (PL) and pyridoxine 5'-phosphate (PNP). Saccharomyces cerevisiae synthesizes PLP via the fungal type de novo PLP synthesis pathway and the PLP salvage pathway.Through these biochemical pathways PLP can be obtained from PL or by synthesis from pyridoxine.
This pathway consists of two steps, in the first pyridoxine 5'-phosphate is synthesized from pyridoxine by the action of pyridoxine kinase enzyme.  The second step is based on the oxidation of pyridoxine 5'-phosphate to pyridoxal phosphate. This reaction requires the participation of oxygen which is reduced to peroxide in this reaction . All these intermediates are present in Figure \ref{bvitam}.

\begin{figure}[t]
\begin{center}
\includegraphics[scale=0.5]{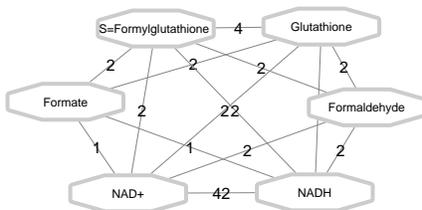}
\caption{Main intermediates involved in removing formaldehyde, the cluster obtained for $k=2$}\label{k2g3}
\end{center}
\end{figure}

Another interesting example, shown in Figure \ref{k2g3},  is the process of removing very reactive and toxic formaldehyde. Although formaldehyde cannot be metabolized from methanol in S. cerevisiae, it can be adopted from plant material or in polluted air and water \cite{yeastDB}. Therefore,  a system for removing this toxic compound should be provided.
The metabolite which plays an important role in this overall defense is glutathione, which has the ability to bind formaldehyde by the spontaneous reaction.
The resulting S-hydroxymethylglutathione is oxidized to S-formyl-glutation with the participation of NAD+ as an oxidizing agent, which is reduced to NADH. By hydrolysis of this compound the products are glutathione and non-toxic formate.

\begin{figure}[t]
\begin{center}
\includegraphics[scale=0.3]{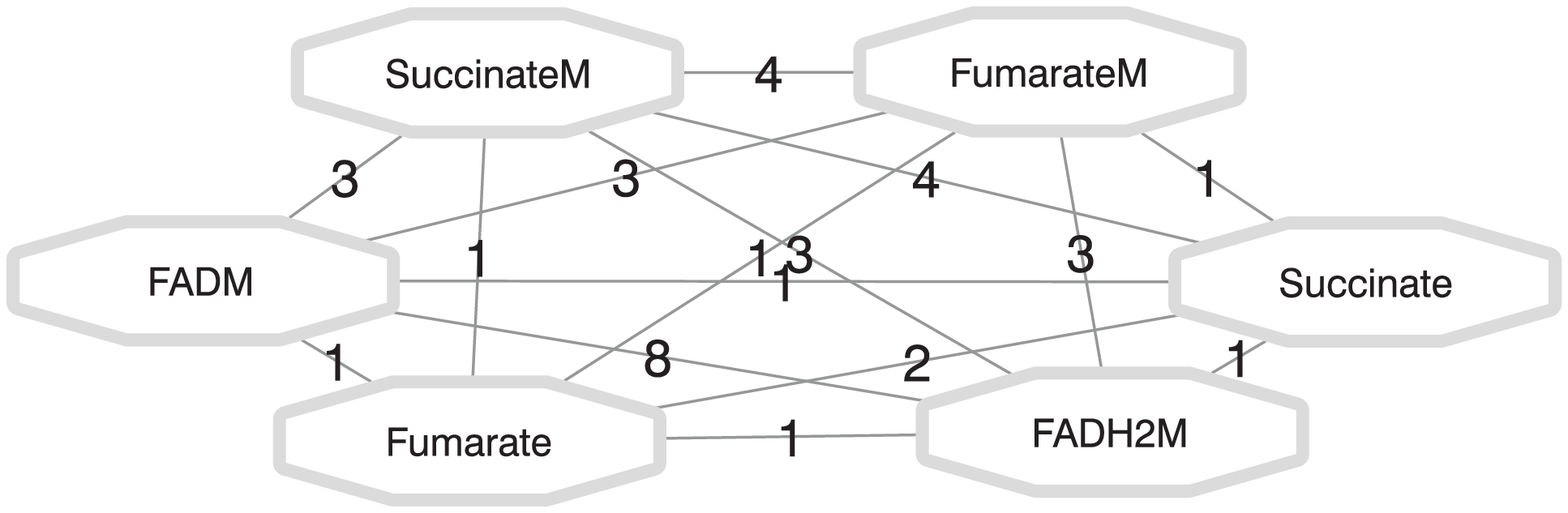}
\caption{The same cluster of oxidation  the succinate to the fumarate, obtained for $k=1,2$ and 3}\label{fumar}
\end{center}
\end{figure}

In the last interesting example, shown in Figure \ref{fumar},  we mention the process of oxidation the succinate to the fumarate. This reaction is possible with the participation of the succinate dehydrogenase enzyme, which is covalently bound to flavin adenine dinucleotide (FAD), which acts as a hydrogen ion acceptor, reducing it to FADH$_2$. For the given reaction, all intermediates are present even in the cluster obtained for $k=1$, so any further relaxation cannot include new members. The algorithm recognized this situation by throwing out the same graph across all three values of $k$.
\subsection{Experimental results obtained on DIMACS instances}

In this subsection we first continue with the comparison with the existing ILP model on a subset of DIMACS instances. In Table \ref{tbl:dimacs} we show the results organized in the same way as for biological instances.  All these instances belong to c-fat, MANN, hamming and johnson families (in Table \ref{tbl:dimacs} the names of instances are shortened). From Table \ref{tbl:dimacs} one can conclude that the proposed VNS achieves 9/9 known optimal solutions. In the rest of 12 cases, VNS achieves 4 best known solutions, while in 8 cases finds new best results.
In regards to the efficiency of the proposed VNS on the larger problem dimensions, the algorithm is also tested on
the challenging set of the rest of 73 DIMACS instances. Although the Max-EkPP is mostly applied on sparse graphs, to achieve completeness of our approach, we decided to test the proposed VNS even on denser DIMACS instances.
For these instances, up to now, no solution is presented in the literature. Although optimallity cannot be proved, small gap values on these instances suggest that VNS obtained high quality solutions. Since the large amount of obtained data will burden the rest of the paper, we decided not to incorporate these results in this paper, but they are available at the first author website http://matinf.pmf.unibl.org/dimacs/.

High quality of the solutions obtained by the proposed VNS clearly indicates that it
is capable for real applications in partitioning biological and other networks in $k$-plexes. The VNS is stable and accurate, while execution times still remain relatively small, even for large instances.

\section{Conclusions}
In this paper, we presented a variable neighborhood search based heuristics for solving the problem of partitioning sparse biological networks into edge-weighted structures called $k$-plexes. This problem, called Max-EkPP problem has been solved for the first time by a metaheuristic approach.  The proposed VNS implements a fast swap-based local search, as well as a specific objective function which favors feasible solutions over infeasible ones, taking into  consideration the degree of every vertex in each partition. An extensive computational analysis is performed on existing sparse biological metabolic networks, as well as on the other artificial instances from literature. From computational point of view, it was shown that the proposed VNS succeeded to achieve all already known optimal or best solutions. It was also able to find new high quality solutions for the other, previously unsolved instances, in a reasonable time.
\begin{table}[t]\caption{Experimental results obtained on smaller and sparser DIMACS instances}
\scriptsize
\centering
\begin{tabular}{llcccccccc}
\hline
$k$&$inst.$&$opt$&$best$&$V_{best}$&$V_{avg}$&$V_{gap}$&$V_t^{tot}$&$ILP$&$ILP_t$\\
\hline
1&c200-1&98711&98711&$opt$&98711&0&234.43&$opt$&47.08\\
2&c200-1&98711&98711&$opt$&98543.2&0.17&202.87&$opt$&567.44\\
3&c200-1&-&98711&$new$&98571.8&0.14&193.7&-&-\\
1&c200-2&213248&213248&$opt$&213246.8&0&540.89&$opt$&0.22\\
2&c200-2&213248&213248&$opt$&212194.6&0.49&360.5&$opt$&47.28\\
3&c200-2&-&213248&$new$&211143.8&0.99&292.97&-&-\\
1&h6-2&65472&65472&$opt$&65472&0&114.53&$opt$&0.2\\
2&h6-2&-&65472&$best$&65472&0&61.91&$best$&$>$10800\\
3&h6-2&-&65472&$best$&65472&0&46.15&$best$&$>$10800\\
1&h6-4&6336&6336&$opt$&6336&0&53.29&$opt$&0.34\\
2&h6-4&-&8184&$new$&8184&0&74.81&6966&$>$10800\\
3&h6-4&-&10560&$new$&10560&0&77.57&4567&$>$10800\\
1&j8-2-4&1260&1260&$opt$&1260&0&7.63&$opt$&0.06\\
2&j8-2-4&-&1365&$new$&1363.5&0.11&10.41&1355&$>$10800\\
3&j8-2-4&-&1996&$best$&1996&0&7.34&$best$&$>$10800\\
1&j8-4-4&-&27874&$new$&27874&0&169.18&27864&$>$10800\\
2&j8-4-4&-&31320&$new$&31147.2&0.55&124.87&12770&$>$10800\\
3&j8-4-4&-&37096&$new$&35910.3&3.2&155.73&12948&$>$10800\\
1&M\_a9&14868&14868&$opt$&14865&0.02&27.55&$opt$&1215.34\\
2&M\_a9&-&23055&$new$&23053.8&0.01&25.96&23047&$>$10800\\
3&M\_a9&33660&33660&$opt$&33660&0&14.23&$opt$&319.24\\
\hline
\end{tabular}\label{tbl:dimacs}
\end{table}

In the deep analysis of the clusters identified  by various values of $k$ on a biological metabolic instance, we confirmed that the algorithm finds many clusters in which the intermediates, that figures in many important metabolic reactions, are highly connected. The relaxation of the adjacency condition leads to obtainment of more useful clusters, which helps in discovering new biological relations or confirming the existing ones.

This research can be extended in several ways. For example, it would be interesting to apply the VNS on solving similar problems, including both biological and non-biological applications. Another direction for the further investigation of this problem can include parallelization of the proposed VNS algorithms and running on some powerful multiprocessor system.

\section*{References}

\bibliography{mybibfile}

\end{document}